\UseRawInputEncoding
\documentclass[mlabstract]{jmlr}

\usepackage{amsmath}
\usepackage{tikz}
\usetikzlibrary{positioning}
\usetikzlibrary{arrows.meta}
\usepackage[english]{babel}
\usepackage{hyphenat}
\usepackage{caption}
\usepackage{mdframed}
\usepackage[font=large,labelfont=bf,labelsep=colon]{caption}
\usepackage{enumitem}

\usepackage{longtable}

\usepackage{booktabs}
\usepackage{siunitx} 

\theorembodyfont{\upshape}
\theoremheaderfont{\scshape}
\theorempostheader{:}
\theoremsep{\newline}

\jmlrvolume{}
\firstpageno{1}
\jmlryear{2025}
\jmlrworkshop{Symmetry and Geometry in Neural Representations}

\title[The Human Brain as a Combinatorial Complex]{The Human Brain as a Combinatorial Complex}

\author{
\Name{Valentina S\'anchez} \Email{v.sanchezmelchor@tilburguniversity.edu}\\
\addr Tilburg University, The Netherlands
\AND
\Name{\c{C}i\c{c}ek G\"uven} \Email{c.guven@tilburguniversity.edu}\\
\addr Tilburg University, The Netherlands
\AND
\Name{Koen Haak} \Email{k.v.haak@tilburguniversity.edu}\\
\addr Tilburg University, The Netherlands
\AND
\Name{Theodore Papamarkou} \Email{theodore@polyshape.com}\\
\addr PolyShape, Greece
\AND
\Name{Gonzalo N\'apoles} \Email{g.r.napoles@tilburguniversity.edu}\\
\addr Tilburg University, The Netherlands
\AND
\Name{Marie \v{S}af\'a\v{r} Postma} \Email{marie.postma@tilburguniversity.edu}\\
\addr Tilburg University, The Netherlands
}

\begin{document}

\maketitle

\begin{abstract}
\sloppy
We propose a framework for constructing combinatorial complexes (CCs) from fMRI time series data that captures both pairwise and higher-order neural interactions through information-theoretic measures, bridging topological deep learning and network neuroscience. Current graph-based representations of brain networks systematically miss the higher-order dependencies that characterize neural complexity, where information processing often involves synergistic interactions that cannot be decomposed into pairwise relationships. Unlike topological lifting approaches that map relational structures into higher-order domains, our method directly constructs CCs from statistical dependencies in the data. Our CCs generalize graphs by incorporating higher-order cells that represent collective dependencies among brain regions, naturally accommodating the multi-scale, hierarchical nature of neural processing. The framework constructs data-driven combinatorial complexes using O-information and S-information measures computed from fMRI signals, preserving both pairwise connections and higher-order cells (e.g., triplets, quadruplets) based on synergistic dependencies. Using NetSim simulations as a controlled proof-of-concept dataset, we demonstrate our CC construction pipeline and show how both pairwise and higher-order dependencies in neural time series can be quantified and represented within a unified structure. This work provides a framework for brain network representation that preserves fundamental higher-order structure invisible to traditional graph methods, and enables the application of topological deep learning (TDL) architectures to neural data.
\end{abstract}

\begin{keywords}
Topological Deep Learning, Combinatorial Complex, Higher-Order Networks, Information Theory, Brain Networks 
\end{keywords}

\section{Introduction}
\label{sec:intro}

The human brain constitutes one of nature's most remarkable examples of organized complexity, exhibiting rich patterns of connectivity and information processing across multiple spatial and temporal scales. As an organized complex system composed of billions of interconnected neurons, the brain demonstrates the defining characteristics of complexity: non-trivial interactions between components that generate emergent properties not reducible to individual parts, and hierarchical organization spanning from molecular to systems-level dynamics \citep{Sporns2022}.

Current approaches to brain network analysis predominantly rely on graph-based representations, where brain regions are modeled as nodes and their pairwise statistical relationships form edges. This framework has become the standard in network neuroscience, enabling fundamental insights into functional connectivity, network topology, and brain organization \citep{Bassett2017,Sporns2010}. The success of this approach has been so profound that neuroscience increasingly needs network science to understand the brain's complex organizational principles \citep{Barabasi2023}. However, graph-based representations face a fundamental limitation: they are inherently restricted to pairwise relationships and cannot capture higher-order interactions—collective dependencies among three or more brain regions that emerge from their joint activity patterns \citep{giusti2016two,Battiston2021,Boccaletti2023}. 

This limitation is particularly problematic for understanding brain function, where information processing often involves synergistic interactions that cannot be decomposed into pairwise relationships. Recent neuroscientific evidence suggests that higher-order interactions may represent important organizing principles of brain networks \citep{Luppi2022,Santoro2024}, motivating our research question: How can we capture higher-order interactions in neural data?

Recent efforts in topological data analysis (TDA) have contributed important tools for studying neural data, including fMRI, by characterizing global shape and connectivity. Most approaches emphasize topological invariants—such as Betti numbers and persistent homology—derived from geometric or distance-based filtrations \citep{Santoro2024}. In contrast, we adopt a data-driven information-theoretic approach without requiring geometric embeddings or distance-based filtrations. We extend graph-theoretical methods beyond pairwise connectivity by constructing combinatorial complexes using multivariate information measures that directly exploit the statistical dependencies in fMRI signals.

From an information-theoretic perspective, these higher-order dependencies can be formally characterized using measures that distinguish between redundant and synergistic information \citep{Varley2025}. Redundant information refers to information that is present in multiple source variables simultaneously, while synergistic information is information that is only accessible when multiple source variables are observed together and cannot be obtained from any individual source. Recent developments in multivariate information theory provide the mathematical foundation for identifying and quantifying these higher-order effects in neural data \citep{Rosas2024,Mediano2022}.

We propose that recent advances in Topological Deep Learning (TDL) offer a promising solution for representing the now quantifiable higher-order interactions in neural time series data. With information-theoretic measures enabling systematic quantification of synergistic dependencies, we can construct combinatorial complexes (CCs)—mathematical structures that generalize graphs to include higher-order relationships \citep{Hajij2022,Papillon2023}. CCs are particularly well-suited for this application because their flexible structure allows direct integration of continuous information-theoretic measures into discrete topological representations, representing both pairwise and higher-order interactions without the closure or inclusion constraints required by other topological structures.

Our data-driven approach generates higher-order cells directly from multivariate dependencies in neural time series. The resulting combinatorial complexes can provide structured representations for combinatorial complex neural networks (CCNNs), which could perform message passing across multiple topological ranks. This would preserve both local pairwise structure and global higher-order patterns, potentially enabling improved learning tasks such as brain state classification or cognitive task decoding.

 We present this work as a first-generation approach that establishes foundational principles while acknowledging current challenges that require future methodological development. This framework bridges topological deep learning and network neuroscience not via topological lifting from lower-order relational scaffolds (e.g., graphs), but through data-driven construction of combinatorial complexes from multivariate statistical dependencies.

\section{From Brain Graphs to Combinatorial Complexes}
\label{sec:methodology}

\subsection{Mathematical Framework}

Traditional brain network analysis constructs an undirected graph $G = (V, E)$, where nodes represent brain regions and edges encode pairwise statistical dependencies. These dependencies are typically inferred from region-wise time series using statistical measures such as Pearson correlation, partial correlation, mutual information, or coherence \citep{Smith2011}. This representation captures at most $\binom{N}{2}$ pairwise relationships, typically stored in an $N \times N$ adjacency matrix.

Combinatorial complexes (CCs) generalize graph-based representations by formally incorporating higher-order cells. In our framework, CCs store both pairwise interactions (edges) and higher-order relationships (e.g., triplets) in a unified, hierarchical structure. The full CC consists of a structured collection of subsets of varying sizes, with up to $\sum_{k=0}^{n} \binom{N}{k+1}$ potential cells. This makes the CC a discrete, higher-dimensional object that grows combinatorially with the number of nodes, posing significant computational challenges common across higher-order network modeling approaches. 

To identify and filter higher-order dependencies, we turn to information-theoretic measures. While several options exist—including mutual information, Partial Information Decomposition (PID), O-information, and S-information \citep{Varley2025}—we focus on the latter two due to their computational tractability and suitability for capturing synergy and redundancy in multivariate systems. CCs relax the closure and embedding constraints imposed by other topological structures (e.g., simplicial complexes), enabling integration with information-theoretic methods in our pipeline.

A \textbf{combinatorial complex} \citep{Hajij2022} is a triple $(S, \mathcal{X}, \mathrm{rk})$ where $S$ is a finite set, $\mathcal{X} \subseteq \mathcal{P}(S) \setminus \{\emptyset\}$, and $\mathrm{rk}: \mathcal{X} \to \mathbb{Z}_{\geq 0}$ satisfies: (i) $\{s\} \in \mathcal{X}$ for all $s \in S$; and (ii) if $x \subseteq y$ with $x, y \in \mathcal{X}$, then $\mathrm{rk}(x) \leq \mathrm{rk}(y)$.
The rank function induces a hierarchy: rank-0 cells are individual nodes, rank-1 cells are edges, and rank-$k$ cells correspond to $k+1$-tuples (e.g., triplets, quadruplets).

\subsection{Combinatorial Complex Construction}

To construct higher-order cells, we quantify multivariate statistical dependencies using two information-theoretic measures: \textbf{S-information} ($\Sigma$) and \textbf{O-information} ($\Omega$) \citep{Rosas2024, Varley2025}. These measures quantify complementary aspects of multivariate statistical dependence across subsets.

\vspace{0.4em}
\vspace{0.4em}
\noindent\textbf{S-information.}
Given random variables $X_1, \dots, X_n$, S-information quantifies the total statistical interdependence within the system:
\begin{equation}
\Sigma(X) = TC(X) + DTC(X)
\end{equation}
where $TC$ and $DTC$ are total and dual total correlations. Higher $\Sigma$ values indicate stronger multivariate dependencies, including potential higher-order synergistic interactions.

\vspace{0.4em}
\noindent\textbf{O-information.} For random variables $X_1, \ldots, X_n$, the O-information measures the system's overall redundancy–synergy bias:
\begin{equation}
\Omega(X) = TC(X) - DTC(X)
\end{equation}
Negative $\Omega$ indicates net synergy; positive values indicate redundancy. In this proof-of-concept, we compute both measures and apply dual thresholding criteria: $\Sigma$ for overall statistical strength and $\Omega$ to select synergy-dominated structures.

\begin{figure}[!htbp]
\captionsetup{font=scriptsize}
\floatconts
  {fig:cc-visualization}
  {\scriptsize\centering
   \caption{
    Toy combinatorial complex from NetSim neural time series. 
    \textbf{Rank-0} cells are brain regions (nodes); 
    \textbf{Rank-1} edges capture pairwise interactions (mutual information $\geq 0.02$); 
    \textbf{Rank-2} hyperedges denote synergistic triplets (S-information $\geq 0.45$, O-information $\lesssim 0$). 
    Top triplets: (2,3,4) with $S=0.51$, $\Omega=0.06$, and (1,2,3) with $S=0.48$, $\Omega=0.04$. 
    Construction pipeline and implementation details are provided in \hyperref[apd:pipeline]{Appendix~A}.
  }}
  {\includegraphics[width=\textwidth]{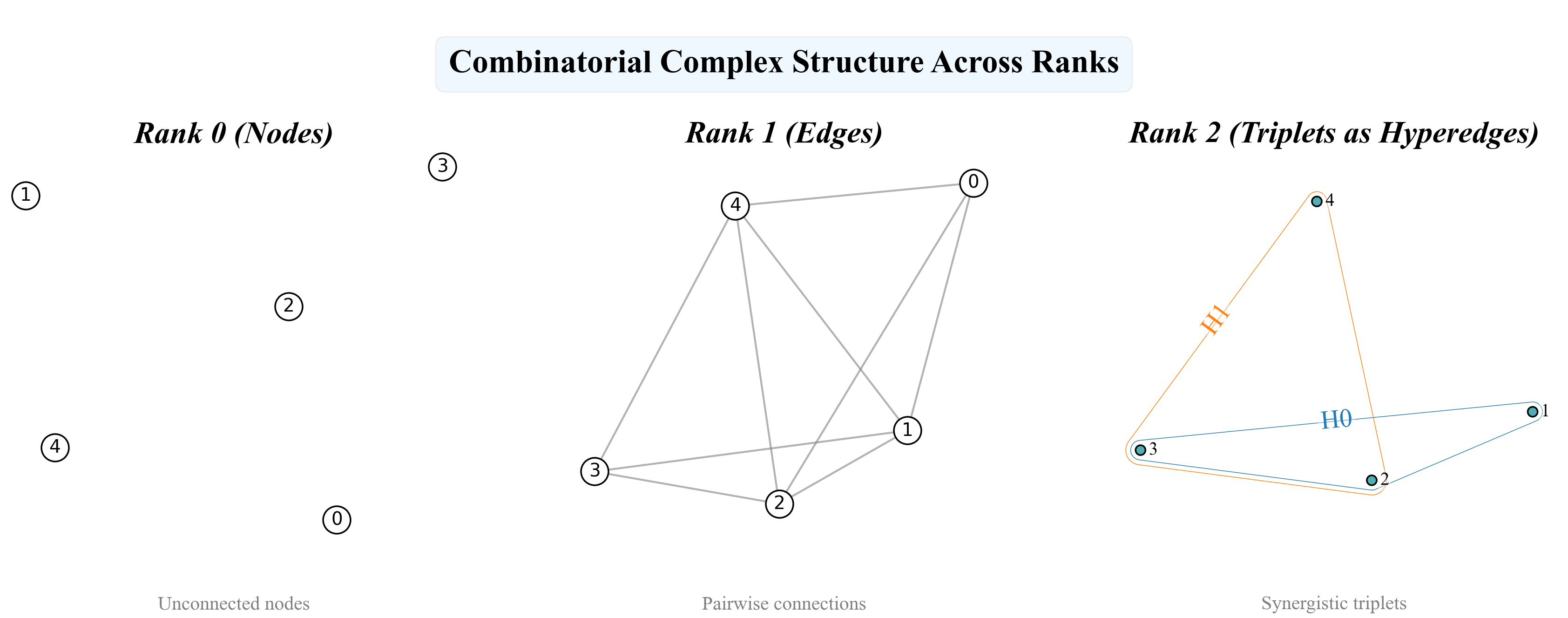}}
\end{figure}

Let $S$ denote a set of variables (e.g., neural time series). For each candidate subset $x \subseteq S$, we compute $\Sigma(x)$ and $\Omega(x)$. Subsets with $\Sigma(x) > \tau$ and $\Omega(x) \lesssim 0$ are retained as rank-$k$ cells in the combinatorial complex, where $k = |x| - 1$. This dual criterion targets synergistic higher-order interactions—collective dependencies that emerge only from joint activity and cannot be decomposed into pairwise relationships. Pairwise dependencies are included as rank-1 cells based on a separate statistical measure (e.g., Pearson correlation).

This yields a combinatorial complex whose structure reflects higher-order informational dependencies inherent in the data (Figure~\ref{fig:cc-visualization}). In contrast to topological filtrations, which rely on geometric constraints, our approach selects cells based on statistical dependencies, uncovering structure intrinsic to the data rather than imposed by geometric priors.

\section{Limitations \& Future directions}
While combinatorial complexes offer a novel way to represent higher-order dependencies in brain data, several limitations remain. The combinatorial growth of candidate cells poses significant computational challenges, which could be mitigated through techniques such as locality-sensitive hashing \citep{Indyk1998LSH}, Kalman filtering \citep{Kalman1960}, and similarity-based preselection. Our current reliance on fixed thresholding may overlook weak but meaningful interactions; future work should explore adaptive criteria, statistical testing, and robust estimators beyond the Gaussian assumption \citep{Varley2025}. Finally, results may depend strongly on how brain regions are defined (e.g., ICA vs. anatomical parcellation \citep{Smith2011}), underscoring the need for standardized benchmarks in higher-order network modeling \citep{bechler2024position}. This framework generalizes to any multivariate time series data and could be integrated with TopoBench \citep{Telyatnikov2025} for topological deep learning applications.

\bibliography{pmlr-sample} 

\clearpage
\appendix
\section{Construction Pipeline And Implementation Details}
\label{apd:pipeline}

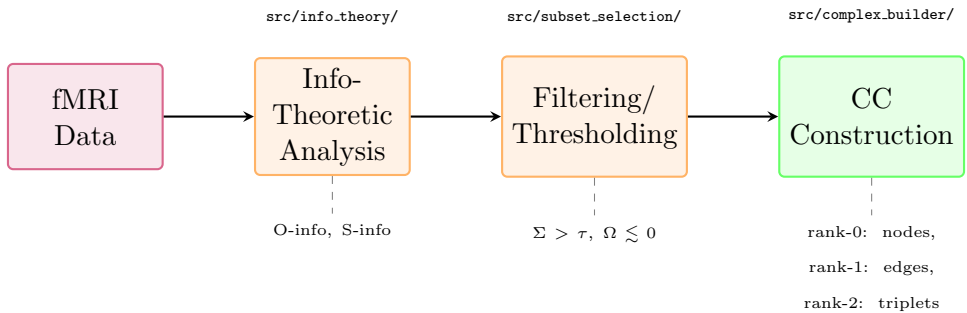
\begin{figure}[htbp]
\captionsetup{font=scriptsize} 
\floatconts
  {fig:pipeline}
  {\scriptsize\centering
   \caption{
    Pipeline for constructing combinatorial complexes (CCs) from fMRI data.
  }}
  {\begin{tikzpicture}[
    node distance = 0.8cm and 1.2cm,
    data/.style = {
        rectangle, 
        draw=purple!60, 
        fill=purple!10, 
        thick,
        text width=1.8cm, 
        text centered, 
        minimum height=1.4cm,
        rounded corners=2pt
    },
    process/.style = {
        rectangle, 
        draw=orange!60, 
        fill=orange!10, 
        thick,
        text width=1.8cm, 
        text centered, 
        minimum height=1.4cm,
        rounded corners=2pt
    },
    processlarge/.style = {
        rectangle, 
        draw=orange!60, 
        fill=orange!10, 
        thick,
        text width=2.2cm, 
        text centered, 
        minimum height=1.6cm,
        rounded corners=2pt
    },
    output/.style = {
        rectangle, 
        draw=green!60, 
        fill=green!10, 
        thick,
        text width=2.2cm, 
        text centered, 
        minimum height=1.6cm,
        rounded corners=2pt
    },
    arrow/.style = {->, >=stealth, thick, black}
]

\node[data] (input) {fMRI Data};
\node[process, right=of input] (analysis) {Info-\\Theoretic\\Analysis};
\node[processlarge, right=of analysis] (filtering) {Filtering/\\Thresholding};
\node[output, right=of filtering] (construction) {CC\\Construction};

\node[below=0.5cm of analysis, text width=1.8cm, align=center] (analysis_detail) {{\tiny O-info, S-info}};
\node[below=0.5cm of filtering, text width=2.2cm, align=center] (filtering_detail) {{\tiny $\Sigma > \tau$, $\Omega \lesssim 0$}};
\node[below=0.5cm of construction, text width=2.2cm, align=center] (construction_detail) {{\tiny rank-0: nodes, rank-1: edges, rank-2: triplets}};

\draw[arrow] (input) -- (analysis);
\draw[arrow] (analysis) -- (filtering);
\draw[arrow] (filtering) -- (construction);

\draw[dashed, gray] (analysis) -- (analysis_detail);
\draw[dashed, gray] (filtering) -- (filtering_detail);
\draw[dashed, gray] (construction) -- (construction_detail);

\node[above=0.3cm of analysis, align=center] {\tiny \texttt{src/info\_theory/}};
\node[above=0.3cm of filtering, align=center] {\tiny \texttt{src/subset\_selection/}};
\node[above=0.3cm of construction, align=center] {\tiny \texttt{src/complex\_builder/}};

\end{tikzpicture}}
\end{figure}

This appendix describes the implementation used to construct combinatorial complexes (CCs) from NetSim \citep{Smith2011} synthetic data and to generate the results shown in Figure~\ref{fig:cc-visualization}, with the overall workflow summarised in Figure~\ref{fig:pipeline}. Our approach focuses on information-theoretic analysis of higher-order dependencies using established tools, with future extensions aimed at applying CC assembly to larger real-fMRI datasets and integrating the framework with topological deep learning methods.

We use NetSim synthetic BOLD time series that simulate realistic fMRI dynamics via a dynamic causal modelling (DCM) neural process coupled to a nonlinear balloon–Windkessel haemodynamic forward model. Each dataset consists of an $N \times T$ matrix $\mathbf{X}$ representing $N$ brain regions over $T$ time points, with neural activity convolved through region-specific haemodynamic response functions, realistic inter-regional HRF variability, and additive Gaussian noise at the BOLD level. While NetSim provides a graph-level ground truth, we use only the generated time series, making it a small, well-controlled, and interpretable yet physiologically informed testbed for proof-of-concept higher-order network construction.

\subsection{Information-theoretic analysis}
\label{app:info-theory}

The core contribution of our pipeline (illustrated in Figure~\ref{fig:pipeline}) lies in integrating higher-order information-theoretic analysis with combinatorial complex (CC) construction. We use the Java Information Dynamics Toolkit (JIDT) \citep{lizier2014jidt}, interfaced via Python using JPype.

\paragraph{Higher-order measures.} For each candidate triplet $(i, j, k)$, we compute S-information and O-information as described in the main text. These quantify the strength of statistical interdependence and the net redundancy–synergy bias within the triplet.

The JIDT library provides robust estimators for multivariate information measures, handling the computational complexity of higher-order statistics. Our wrapper functions in \texttt{src/info\_theory/jidt\_interface/} standardize the input/output format and provide error handling for the JPype bridge.

\subsection{Combinatorial complex construction framework}
\label{app:cc-construction}

\paragraph{Rank assignment.}  
Brain regions are treated as rank-0 cells (nodes). Rank-1 cells (edges) are added based on pairwise mutual information (MI), using a fixed threshold of 0.02 to ensure sparsity while retaining meaningful connections. For rank-2 cells (triplets), we compute both S-information ($\Sigma$) and O-information ($\Omega$) using Gaussian estimators from JIDT \citep{lizier2014jidt}. A triplet is retained as a rank-2 cell if both its $\Sigma$ value exceeds 0.45 and $\Omega \lesssim 0$, reflecting strong multivariate dependencies with synergy-dominated character. This dual criterion ensures we capture higher-order interactions that are both informationally rich and synergistic rather than redundant. Although our current implementation focuses on ranks 0–2 for tractability, the framework generalizes naturally to higher-rank structures.

\paragraph{Thresholding strategy.}  
The choice of fixed thresholds was guided by exploratory runs on the NetSim dataset. $\Sigma$ is used to quantify the overall strength of statistical dependencies within a triplet, while $\Omega$ is computed in parallel to diagnose whether synergy or redundancy dominates. Together, they allow us to detect strongly coupled subsets and interpret the nature of their higher-order interactions. In this proof-of-concept, we implement dual selection criteria: $\Sigma > 0.45$ identifies subsets with strong statistical dependencies, while $\Omega \lesssim 0$ ensures we retain only synergy-dominated structures. We use “$\lesssim 0$” rather than “$< 0$” to accommodate weak synergy-dominant structures with small positive $\Omega$ values. This dual thresholding approach filters for both statistical strength and synergistic character. As emphasized by \citet{Varley2025}, fixed thresholds can miss weak but relevant interactions, motivating future integration of adaptive criteria, statistical tests, and more robust estimators.

\paragraph{Illustrative example and interpretation.}  
Figure~\ref{fig:cc-visualization} shows the CC constructed from subject 1 in the \texttt{sim1.mat} NetSim dataset (50 subjects, 5 regions, 200 time points). We selected this example for its simplicity as a toy testbed. The top two triplets—(2,3,4) with $\Sigma = 0.51$, $\Omega = 0.06$, and (1,2,3) with $\Sigma = 0.49$, $\Omega = 0.04$—illustrate how our method captures structured higher-order dependencies that are invisible to graph-based representations. While $\Sigma$ reflects the overall strength of multivariate dependencies, $\Omega$ helps reveal the balance between synergy and redundancy: the mildly positive values here suggest a mixed interaction pattern with a synergy-dominant core. In larger or more complex datasets, $\Omega$ will further aid in distinguishing interaction motifs. Even in this minimal setting, the resulting CC reveals triadic structures that reflect non-pairwise organization in the neural signals.

\subsection{Computational considerations and software stack}
\label{app:computational}

\paragraph{Scalability.} 
The computational cost of constructing combinatorial complexes grows combinatorially with rank.  
For a dataset with $N$ regions, the number of candidate rank-$k$ cells is $\binom{N}{k+1}$, and evaluating each requires computing information-theoretic quantities over $(k+1)$ variables.  
For example:
\begin{itemize}
    \item Rank-1 (edges): $O(N^2)$
    \item Rank-2 (triplets): $O(N^3)$
    \item Rank-$(q-1)$: $O(N^q)$
\end{itemize}
In practice, our proof-of-concept restricts analysis to triplets for tractability.  
With typical NetSim dimensions ($N \approx 5$, $T \approx 200$), the pipeline executes efficiently, requiring evaluation of only $\binom{5}{3} = 10$ triplets.
However, the combinatorial growth becomes prohibitive for realistic brain parcellations: standard atlases with $N = 100$ regions would require evaluating $\binom{100}{3} = 161,700$ triplets.
Beyond $N \approx 50$, computational requirements exceed typical desktop capabilities without sophisticated optimization.
The modular design in \texttt{src/info\_theory/} supports future scalability improvements via candidate pre-filtering, parallelisation, and distributed computation, and could incorporate advanced strategies such as locality-sensitive hashing \citep{Indyk1998LSH}, Kalman filtering~\citep{Kalman1960}, and similarity-based preselection to further reduce the search space.

\paragraph{Software dependencies.}\mbox{}\\

The implementation builds on:
\begin{itemize}
    \item \textbf{Core computation:} NumPy, SciPy for numerical operations
    \item \textbf{JIDT interface:} JPype1 for Java–Python bridging
    \item \textbf{Visualization:} Matplotlib for time series plots, HyperNetX for combinatorial complex visualisation (Figure~\ref{fig:cc-visualization})
\end{itemize}

\subsection{Reproducibility and code availability}
\label{app:reproducibility}

The public implementation and scripts to reproduce the figure are available at \href{https://github.com/ValentinaSanchezMelchor/TopoBrainX/tree/main}{\texttt{TopoBrainX}}.

\paragraph{Future extensions.} The current implementation establishes a foundation for constructing combinatorial complexes directly from time series data. While this framework was designed with fMRI in mind, it generalizes to any multivariate time series or signal data. In this proof-of-concept, we focus on demonstrating the full pipeline rather than on exploratory analysis or interpretation. Planned developments include scaling to large datasets while mitigating combinatorial explosion, applying the method to real fMRI data for data analysis and interpretation, incorporating adaptive thresholding strategies, and integrating with TopoBench \citep{Telyatnikov2025} for downstream topological deep learning applications.

\end{document}